\documentclass[prd,twocolumn,showpacs,floatfix,amsmath,amssymb,floatfix]{revtex4}
\usepackage{graphicx,color,dcolumn,booktabs,bm}
\usepackage{longtable,lscape}
\usepackage{txfonts}
\usepackage{overpic}
\usepackage{amssymb}
\usepackage{indentfirst}
\usepackage{feynmf}   %{feynmp}
\usepackage{slashed}  %for Feynman symbols
\usepackage{cases}
\usepackage{color}
\usepackage{multirow}
\usepackage{epstopdf}
\usepackage{graphicx,color,dcolumn,booktabs,bm}

\graphicspath{{Figures/}} %
\usepackage[colorlinks,
                      citecolor=blue,
                      anchorcolor=red,
                      menucolor=red,
                      linkcolor=red,
                      filecolor=red,
                      runcolor=red,
                      urlcolor=blue,
                      frenchlinks=red]{hyperref}
\begin{document}

\title{Hidden Charm Decays of $Y(4626)$ in a $D_{s}^{*+}{D}_{s1}(2536)^-$ Molecular Frame}

\author{Zi-Li Yue$^{1}$}\email{yuezili@seu.edu.cn}
\author{Yue Pan$^{1}$}\email{panyue@seu.edu.cn}
\author{Dian-Yong Chen$^{1,2}$}\email{chendy@seu.edu.cn}
\affiliation{
 $^{1}$ School of Physics, Southeast University,  Nanjing 210094, China\\
 $^2$Lanzhou Center for Theoretical Physics, Lanzhou University, Lanzhou 730000, China}

\begin{abstract}
In this work, we investigate the hidden charm decays properties of $Y(4626)$, where $Y(4626)$ is assigned as a $S-$wave $D_{s}^{*+}D_{s1}(2536)^{-}$ molecular state with $J^{PC}=1^{--}$. The partial widths of the processes $Y(4626)\to J/\psi\eta$, $J/\psi\eta^{\prime}$, $\eta_{c}\phi$, and $ \chi_{cJ}\phi,\ (J=\{0,1,2\})$ are estimated by employing the effective Lagrangian approach. The present estimations indicate that the partial widths of the $J/\psi\eta$ and $J/\psi \eta^\prime$ channels are of the order of 1 MeV, while the one of $\chi_{c1}\phi$ is of the order of 0.1 MeV. Thus, we propose to further examine the molecular interpretation of $Y(4626)$ by searching it in the cross sections for the $e^{+}e^{-}\to J/\psi\eta^{(\prime)}$ processes, which should be accessible by the BES \uppercase\expandafter{\romannumeral 3} and Belle \uppercase\expandafter{\romannumeral 2}.
\end{abstract}

\pacs{ }

\maketitle

%%%%%%%%%%%%%%%%%%%%%%%%%%%%%%%%%%
\section{Introduction}
\label{sec:introduction}
Numerous charmonium-like states, referred to as $XYZ$ states~\cite{BESIII:2013ris,BESIII:2013ouc,CDF:2009jgo,BaBar:2006ait,Belle:2007umv,Belle:2014nuw,D0:2013jvp,CMS:2013jru,LHCb:2016axx,LHCb:2021uow,BaBar:2012hpr}, have been observed following the discovery of $X(3872)$ in the year $2003$~\cite{Belle:2003nnu}. Among charmonium-like states, a multitude of vector states with $J^{PC}=1^{--}$, typically reported in the $e^+ e^-$ annihilation processes, are usually denoted as $Y$ states. As a typical example of the $Y$ state family, $Y(4260)$ was first observed by the BaBar Collaboration in the cross sections of $e^+e^- \to \pi^{+}\pi^{-}J/\psi$ by using the initial-state radiation technique in 2005~\cite{BaBar:2005hhc}. Subsequently, the CLEO~\cite{CLEO:2006tct}, Belle~\cite{Belle:2013yex,Belle:2007dxy}, BaBar~\cite{BaBar:2012vyb} and BES III~\cite{BESIII:2020oph, BESIII:2016bnd} Collaborations independently confirmed the existence of $Y(4260)$ through the same process. The mass and width of PDG average are~\cite{ParticleDataGroup:2022pth}, 
\begin{eqnarray}
m_{Y(4260)}&=&\left(4222.5\pm2.4\right)~\mathrm{MeV},\nonumber \\
\Gamma_{Y(4260)} &=&	\left(48\pm8\right)~\mathrm{MeV},
\end{eqnarray}
respectively. As indicted in Ref.~\cite{Chen:2010nv,Chen:2015bft}, categorizing $Y(4260)$ into $\psi$ family becomes questionable due to its exhibited properties as non-$q\bar{q}$ state. The observed mass of $Y(4260)$ lies approximately $70~\mathrm{MeV}$ below the $D\bar{D}_{1}(2420)$ threshold. Thus, the $D\bar{D}_{1}(2420)$ molecular state interpretation for $Y(4260)$ have been proposed, and various approach, such as the effective Lagrangian approach~\cite{Chen:2016byt}, chiral quark model~\cite{Li:2013bca}, the Lattice QCD~\cite{Chiu:2005ey}, potential model \cite{Ding:2008gr, Close:2009ag, Close:2010wq, Dong:2019ofp} and Bethe-Salpeter formalism~\cite{Wang:2020lua}, have been utilized to explore the possibility of interpreting the $Y(4260)$ as a $D\bar{D}_{1}(2420)$ molecular state.

In the same energy range, there exists another $Y$ state known as $Y(4360)$, which was observed in the cross section for the process $e^{+}e^{-}\to\pi^{+}\pi^{-}\psi(2S)$ near $4.32~\mathrm{GeV}$ by the Babar and Belle Collaborations~\cite{BaBar:2006ait,Belle:2007umv,BaBar:2012hpr}. The resonance parameters of the $Y(4360)$ were measured to be~\cite{ParticleDataGroup:2022pth}, 
\begin{eqnarray}
	m_{Y(4360)}=(4374\pm7)~\mathrm{MeV},\nonumber\\
	\Gamma_{Y(4360)}=(118\pm12)~\mathrm{MeV},
\end{eqnarray}
respectively. Similarly to the case of $Y(4260)$, $Y(4360)$ cannot be unambiguously assigned as a conventional charmonium, as discussed in Ref.~\cite{Chen:2013bha,Chen:2015bft}. In addition, the mass of $Y(4360)$ lies approximately $60$ MeV below the threshold of $D^{\ast}\bar{D}_{1}(2420)$. In~\cite{Close:2009ag, Close:2010wq} a deeply $D^\ast \bar{D}_1(2430)$ bound state, which may corresponding to $Y(4260)$ or $Y(4360)$, was found with the $\pi$ exchange interaction. Besides, the interactions between a pair of charmed mesons in the $T-$doublet were investigated systematically in Ref.~\cite{Wang:2021ajy}.

The story about the vector charmonium-like states goes on. In the year $2019$, the Belle Collaboration reported the observation of $Y(4626)$ in the cross sections for the process $e^{+}e^{-}\to D_{s}^{+}D_{s1}(2536)^{-}$  with a significance of $5.9\sigma$~\cite{Belle:2019qoi}. The mass and width of $Y(4626)$ were reported to be be~\cite{Belle:2019qoi},
\begin{eqnarray}
m_{Y(4626)}&=& \left(4629.5^{+6.2}_{-6.0}(\mathrm{stat}.)\pm0.4(\mathrm{syst.}) \right) \ \mathrm{MeV}, \nonumber\\
\Gamma_{Y(4626)}&=&\left(49.8^{+13.9}_{-11.5}(\mathrm{stat}.)\pm4.0(\mathrm{syst.})\right) \ \mathrm{MeV},
\end{eqnarray}
respectively. Shortly after, the Belle Collaboration observed a similar  structure with a $3.4\sigma$ significance in the invariant mass spectrum of $D_{s}^{+}D_{s2}^{*}(2573)^{-}$ using the initial-state radiation technique in the $e^{+}e^{-}$ annihilation~\cite{Belle:2020wtd}. The mass and decay width were determined to be $\left(4619.8_{-8.0}^{+8.9}(\mathrm{stat}.)\pm2.3(\mathrm{syst}.)\right)$ $\mathrm{MeV}$ and $\left(47.0^{+31.3}_{-14.8}(\mathrm{stat}.)\pm4.6(\mathrm{syst}.)\right)$ $\mathrm{MeV}$, respectively. Interestingly,  $Y(4626)$ is close to previously observed $Y(4630)$~\cite{Belle:2008xmh} in the $\Lambda_{c}^{+}\Lambda_{c}^{-}$ invariant mass distribution and $Y(4660)$ observed in the process $e^{+}e^{-}\to \pi^{+}\pi^{-}\psi(2S)$~\cite{Belle:2007umv}. These discoveries contribute to the complexity of the energy region under investigation.
These states exhibit consistent masses, decay widths, and quantum numbers within the measured uncertainties, suggesting a possible common underlying structure. As a result, there has been growing interest in their nature and potential explanations. For instance, in Ref. ~\cite{Bugg:2008sk}, the authors proposed that a form factor of reasonable radius of interaction could explain the mass shift between $Y(4630)$ and $Y(4660)$, and in the $\psi^\prime f_0(980)$ molecular picture taking into account $\Lambda_c^+ \bar{\Lambda}_c^-$ final state interaction could explain these two structures~\cite{Guo:2010tk}. Motivated by the proximity of the $Y(4630)$ mass to the $\Lambda_{c}\bar{\Lambda}_{c}$ threshold, the authors in Ref.~\cite{Song:2022yfr}, proposed $Y(4630)$ to be a $\Lambda_{c}\bar{\Lambda}_{c}$ molecular state. Beyond molecular interpretations, the properties of $Y(4630)$ have been investigated within the tetraquark frame. In Ref.~\cite{Cotugno:2009ys}, $Y(4630)$ was considered as the first radial excitation of the $\ell=1$ state of the $[cd][\bar{c}\bar{d}]$ diquark-antidiquark bound state, while the estimations using the QCD sum rule~\cite{Zhang:2020gtx} and the multiquark color flux tube model~\cite{Deng:2019dbg} indicated that this state could be a $P$-wave $[cs][\bar{c}\bar{s}]$ tetraquark state. Moreover, attempts have been made to categorize these charmonium-like states within the conventional charmonium framework. For instance, $Y(4660)$ was interpreted as a good candidate for the $5^{3}S_{1}$ charmonium state in Ref.~\cite{Ding:2007rg}, and the authors in Ref.~\cite{Wang:2020prx} investigated the higher charmonium mass spectrum using the unquenched potential model, suggesting that $Y(4626)$, $Y(4630)$ and $Y(4660)$ may be the mixtures of $6S$ and $5D$ $c\bar{c}$ states. 
. 

It should be noted that the mass of $Y(4626)$ is close to the $D_{s}^{*+}{D}_{s1}(2536)^{-}$ threshold. Furthermore, the mass difference between $D_{s}^{*+}{D}_{s1}(2536)^{-}$ and $D^{*}\bar{D}_{1}(2420)$ is almost identical to the mass difference between $Y(4626)$ and $Y(4390)$, i.e.,
\begin{eqnarray}
m_{Y(4626)}-m_{Y(4390)}\approx (m_{D_s^{*}}+m_{\bar{D}_{s1}})-(m_{D^{*}}+m_{\bar{D}_{1}}),
\end{eqnarray}
Previous study in Ref.~\cite{Chen:2017abq} identified the $Y(4390)$ as a molecular state consisting of $D^{*}\bar{D}_{1}(2420)$, and its hidden-charm decays were investigated. Consequently, it is reasonable to consider the $Y(4626)$ as an $S$-wave $D_{s}^{*+}{D}_{s1}(2536)^{-}$ molecular state based on SU(3) symmetry. The quasipotential Bethe-Salpeter equation calculations with one-boson-exchange model suggested $Y(4626)$ as a $D_s^{\ast+} D_{s1}(2536)^-$ molecular state with $J^{PC}=1^{--}$~\cite{He:2019csk}, and such molecular interpretation was also supported by the  estimations based on heavy-quark spin and SU(3)-flavor symmetries~\cite{Peng:2022nrj}. Along this way, in the present work, we further examine the plausibility of the $D_{s}^{*+}D_{s1}(2536)^{-}$ molecular interpretation to $Y(4626)$ by investigating the hidden charm decay behaviors of $Y(4626)$ with an effective Lagrangian approach, which may provide some useful information for further observations of $Y(4626)$ by the BES III and Belle II Collaborations in future.

This work is organized as follows. After the introduction, the hadronic  molecule structure of $Y(4626)$ is discussed in \ref{sec:Sec2}. The hidden charm decays including $Y(4626)\to J/\psi\eta$, $Y(4626)\to J/\psi\eta^{\prime}$, $Y(4626)\to \eta_{c}\phi$, $Y(4626)\to \chi_{c0}\phi$, $Y(4626)\to \chi_{c1}\phi$ and $Y(4626)\to \chi_{c2}\phi$ are estimated in \ref{sec:Sec3}. The numerical results and related discussions are presented in \ref{sec:Sec4}, and a short summary is provided in \ref{sec:Sec5}.

\section{Hadronic molecular structures of the $Y(4626)$}
\label{sec:Sec2}
In the current study, the $Y(4626)$ is assigned as an $S$-wave $D_{s}^{*+}D_{s1}(2536)^{-}$ hadronic molecule with  $I(J^{PC})=0(1^{--})$. Given that the isospin of the state is $0$ and its spin parity $J^{P}$ is determined in the partial wave decomposition, we only need to give a flavor function for the $Y(4626)$ with $C=-1$,
\begin{eqnarray}
\left|D_{s}^{*}\bar{D}_{s1}\right\rangle=\frac{1}{\sqrt{2}}\Bigg[\left|D_{s}^{*+}D_{s1}^{-}\right\rangle-c\left|D_{s}^{*-}D_{s1}^{+}\right\rangle \Bigg]\ ,
\end{eqnarray}
here, we use the appointment $\mathcal{C}D_{s}^{\pm}\mathcal{C}^{-1}=D_{s}^{\mp}$, $\mathcal{C} D_{s1}^{\pm}\mathcal{C}^{-1}=D_{s1}^{\mp}$, and $\mathcal{C} D_{s}^{*\pm}\mathcal{C}^{-1}=-D_{s}^{*\mp}$. With these relations, it is straightforward to obtain $c=-1$ for $Y(4626)$~\cite{He:2019csk,Liu:2008fh,Wang:2020dya,Liu:2009qhy,Sun:2012sy}.

In the present estimations, we employ an effective Lagrangian approach to describe the interaction between  $Y(4626)$ and its components, which is,
\begin{eqnarray}
\mathcal{L}_Y&=&g_{Y}\varepsilon_{\mu\nu\alpha\beta}\partial^{\mu}Y^{\nu}(x)\int dy\Phi\left(y^{2}\right)\nonumber\\&\times&D_{s}^{*+\alpha }\left(x+\omega_{D_{s1}^{-}}y \right)D_{s1}^{-\beta }\left(x-\omega_{D_{s}^{*+}}y\right)+\mathrm{c.c.},\label{Eq:1}
\end{eqnarray}
with $\omega_{D_{s1}^{-}}=m_{D_{s1}^{-}}/(m_{D_{s1}^{-}}+m_{D_{s}^{*+}})$ and $\omega_{m_{D_{s}^{*+}}}=m_{D_{s}^{*+}}/(m_{D_{s1}^{-}}+m_{D_{s}^{*+}})$. $\Phi(y^{2})$ is the correlation function introduced to describe the interior structure of $Y(4626)$, and it Fourier transformation is,
\begin{eqnarray}
\Phi\left(y^{2} \right)=\int \frac{d^{4}p}{(2\pi)^{4}}e^{-ipy}\tilde{\Phi}\left( -p^{2} \right).
\end{eqnarray}
We adopt the Gaussian form $\tilde{\Phi}(-p^{2})$ \cite{Faessler:2007gv,Faessler:2007us,Xiao:2020alj,Xiao:2019mvs,Chen:2015igx,Xiao:2016hoa} to ensure that the correlation function drops rapidly enough in the ultraviolet region of Euclidean space and depicts the molecular inner configuration,
\begin{eqnarray}
\tilde{\Phi}\left(p_{E}^{2} \right)=\mathrm{exp}\left(-p_{E}^{2}/\Lambda_{Y}^{2}\right),
\end{eqnarray}
where $\Lambda_{Y}$ is the size parameter, which characterizes the distribution of the molecular constituents.

\begin{figure}[t]
  \centering
  \includegraphics[width=7cm]{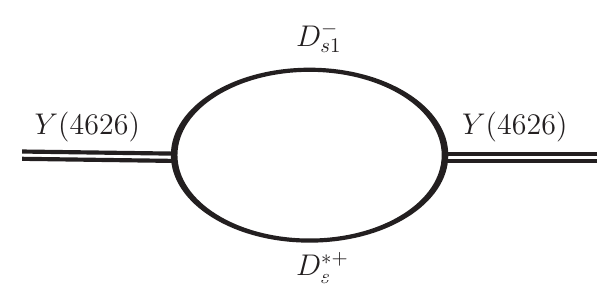}
  \caption{The mass operator of the $Y(4626)$ within the $D_s^{\ast+}+D_{s1}(2536)^-$ molecular frame.}\label{Fig:Tri1}
\end{figure}

The coupling constant $g_{Y}$ can be determined by the compositeness condition \cite{Weinberg:1962hj,Salam:1962ap,vanKolck:2022lqz,Xiao:2020ltm,Chen:2015igx}, which is,
\begin{eqnarray}
Z=1-\Sigma^{\prime}\left(m_{Y}^{2}\right),\label{Eq:2}
\end{eqnarray}
generally, the renormalization constant $Z$ ranges from $0$ to $1$. When $Z=0$, $Y(4626)$ is described as a pure bound state. The mass operator $\Sigma^{\mu\nu}_{Y}$ can be split into the transverse part $\Sigma_{Y}$ and the longitudinal part $\Sigma_{Y}^{L}$,
\begin{eqnarray}
\Sigma^{\mu\nu}_{Y}\left(p^2\right)=g^{\mu\nu}_{\bot}\Sigma_{Y}\left(p^2\right)+\frac{p^{\mu}p^{\nu}}{p^{2}}\Sigma_{Y}^{L}\left(p^2\right),
\end{eqnarray}
with $g^{\mu\nu}_{\bot}=g^{\mu\nu}-p^{\mu}p^{\nu}/p^{2}$ and $g^{\mu\nu}_{\bot}p_{\mu}=0$.

With the effective Lagrangian given in Eq.~(\ref{Eq:1}), the concrete form of the mass operator corresponding to Fig.~\ref{Fig:Tri1} of $Y(4626)$ can be written as,
\begin{eqnarray}
\Sigma^{\mu\nu}\left(m_{Y}^{2}\right)&=&g_{Y}^{2}\epsilon_{\lambda\nu\alpha\beta}\epsilon_{\rho\mu\theta\tau}(-ip^{\lambda})(ip^{\rho})\nonumber\\&\times&\int \frac{d^{4}q}{(2\pi)^{4}}\tilde{\Phi}^{2}\left[-(q-\omega_{D_{s}^{*}D_{s1}}p),\Lambda^{2}\right]\nonumber\\&\times&\frac{-g^{\beta\tau}+(p^{\beta}-q^{\beta})(p^{\tau}-q^{\tau})/m_{D_{s1}}^{2}}{(p-q)^{2}-m_{D_{s1}}^{2}}\nonumber\\&\times&\frac{-g^{\alpha\theta}+q^{\alpha}q^{\theta}/m_{D_{s}^{*}}^{2}}{q^{2}-m_{D_{s}^{*}}^{2}}.
\end{eqnarray}

\section{Hidden charm decays of $Y(4626)$}
\label{sec:Sec3}

\begin{figure}[t]
\begin{tabular}{cc}
  \centering
  \includegraphics[width=4.2cm]{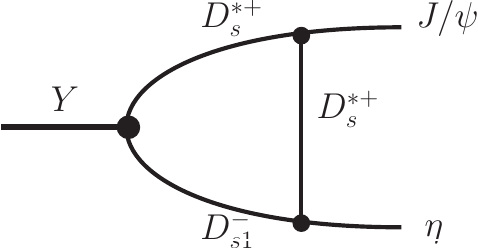}&\hspace{-0.2cm}
 \includegraphics[width=4.2cm]{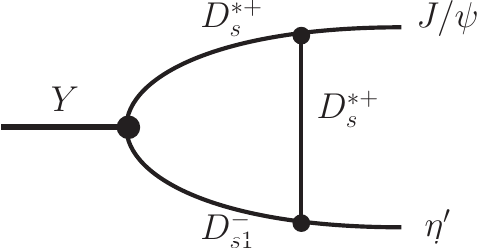}\\
\\
 $(a)$ & $(b)$ \\ \\
  \includegraphics[width=4.2cm]{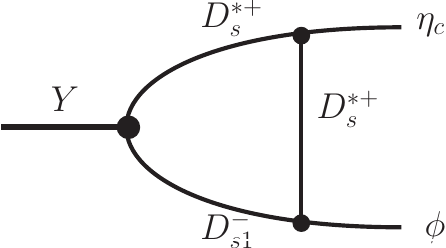}&\hspace{-0.2cm}
 \includegraphics[width=4.2cm]{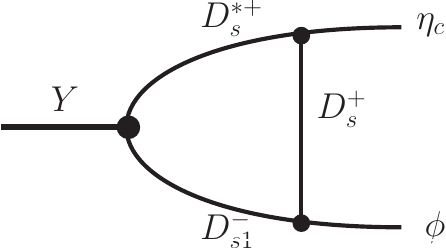}\\
\\$(c)$ & $(d)$ \\ \\
  \includegraphics[width=4.2cm]{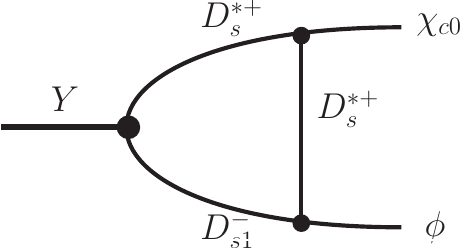}&\hspace{-0.2cm}
 \includegraphics[width=4.2cm]{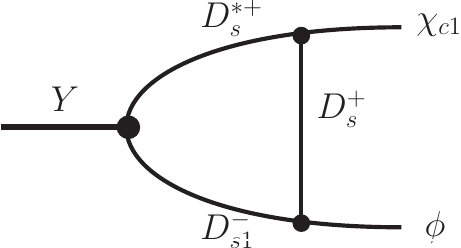}\\
 \\ $(e)$ & $(f)$ \\ \\
\includegraphics[width=4.2cm]{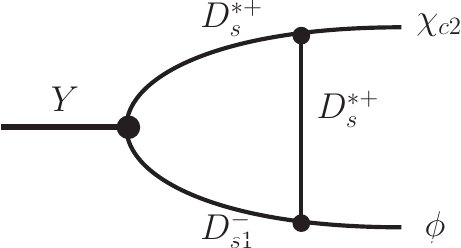}
 \\ \\ $(g)$  \\ 
 \end{tabular}
  \caption{The typical diagrams contributing to 
$Y\to J/\psi\eta$ (diagram (a)), $Y\to J/\psi\eta^{\prime}$ (diagram (b)), $Y\to \eta_{c}\phi$ (diagram (c) and (d)), $Y\to \chi_{c0}\phi$ (diagram (e)), $Y\to \chi_{c1}\phi$ (diagram (f)) and $Y\to \chi_{c2}\phi$ (diagram (g)).}\label{Fig:Tri2}
\end{figure}

In the present work, $Y(4626)$ is considered as a $D_{s}^{*+} {D}_{s1}(2536)^{-}$ molecule with $J^{PC}=1^{--}$. In the molecular frame, the $Y(4626)$ couples to its components $D_{s}^{*+} {D}_{s1}(2536)^{-}+c.c.$ and its components transit into a charmonium and a light meson by exchanging a proper charmed-strange meson. In the present estimations, we select six possible hidden charm decay channels, which are $Y\to J/\psi\eta$, $J/\psi\eta^{\prime}$, $\eta_{c}\phi$, $\chi_{c0}\phi$, $\chi_{c1}\phi$, $\chi_{c2}\phi$. The diagrams contributing to these decay processes at the hadron level are listed in Fig.~\ref{Fig:Tri2}.

\subsection{Effective Lagrangians}

In the present estimations, we employ the effective Lagrangian approach to evaluate the diagrams in Fig.~\ref{Fig:Tri2}. Considering the heavy-quark symmetry and chiral symmetry, the relevant effective Lagrangians can be constructed as~\cite{He:2019csk,Casalbuoni:1996pg,Cheng:1992xi,Yan:1992gz,Wise:1992hn,Huang:2021kfm,Wang:2022jxj,Ding:2008gr},
\begin{eqnarray}
\mathcal{L}_{\psi \mathcal{D}^{(*)}\mathcal{D}^{(*)}}&=&-ig_{\psi \mathcal{D}\mathcal{D}}\psi_{\mu}\mathcal{D}^{\dagger}\stackrel{\leftrightarrow}{\partial}^{\mu}\mathcal{D}\nonumber\\&+&ig_{\psi \mathcal{D}^{*}\mathcal{D}}\epsilon^{\mu\nu\alpha\beta}\partial_{\mu}\psi_{\nu}\left(\mathcal{D}_{\alpha}^{*}\stackrel{\leftrightarrow}{\partial}_{\beta}\mathcal{D}^{\dagger}-\mathcal{D}\stackrel{\leftrightarrow}{\partial}_{\beta}\mathcal{D}_{\alpha}^{*\dagger}\right)\nonumber\\&+&ig_{\psi \mathcal{D}^{*}\mathcal{D}^{*}}\psi^{\mu}\left(\mathcal{D}_{\nu}^{*}\stackrel{\leftrightarrow}{\partial}^{\nu}\mathcal{D}_{\mu}^{*\dagger}+\mathcal{D}_{\mu}^{*}\stackrel{\leftrightarrow}{\partial}^{\nu}\mathcal{D}_{\nu}^{*\dagger}\right.\nonumber\\&-&\left.\mathcal{D}_{\nu}^{*}\stackrel{\leftrightarrow}{\partial}_{\mu}\mathcal{D}^{*\nu \dagger}\right)+\mathrm{H.c.},\nonumber\\
%\end{eqnarray}
%\begin{eqnarray}
\mathcal{L}_{\eta_{c}\mathcal{D}^{*}\mathcal{D}^{(*)}}&=&-ig_{\eta_{c}\mathcal{D}^{*}\mathcal{D}}\eta_{c}\left(\mathcal{D}\stackrel{\leftrightarrow}{\partial}_{\mu}\mathcal{D}^{*\dagger\mu}+\mathcal{D}^{*\mu}\stackrel{\leftrightarrow}{\partial}_{\mu}\mathcal{D}^{\dagger}\right)\nonumber\\&+&ig_{\eta_{c}\mathcal{D}^{*}\mathcal{D}^{*}}\epsilon^{\mu\nu\alpha\beta}\partial_{\mu}\eta_{c}\mathcal{D}^{*}_{\nu}\stackrel{\leftrightarrow}{\partial}_{\alpha}\mathcal{D}_{\beta}^{*\dagger}+\mathrm{H.c.},\nonumber\\
%\end{eqnarray}
%\begin{eqnarray}
\mathcal{L}_{\chi_{cJ}\mathcal{D}^{(*)}\mathcal{D}^{(*)}}&=&g_{\chi_{c0}}\mathcal{D}_{i}\mathcal{D}_{i}^{*\dagger}\nonumber\\&+&g_{\chi_{c0}\mathcal{D}^{*}\mathcal{D}^{*}}\mathcal{D}_{i\mu}^{*}\mathcal{D}_{i}^{*\mu\dagger}+ig_{\chi_{c1}\mathcal{D}^{*}\mathcal{D}}\left(\mathcal{D}^{*}_{i\mu}\mathcal{D}_{i}^{\dagger}-\mathcal{D}_{i}\mathcal{D}_{i\mu}^{*\dagger}\right)\nonumber\\&+&g_{\chi_{c2}\mathcal{D}^{*}\mathcal{D}^{*}}\chi_{c2}^{\mu\nu}\mathcal{D}^{*}_{i\mu}\mathcal{D}^{*\dagger}_{i\nu},\nonumber\\
%\end{eqnarray}
%\begin{eqnarray}
\mathcal{L}_{\mathcal{D}^{*}\mathcal{D}_{1}\mathcal{P}}&=&g_{\mathcal{D}^{*}\mathcal{D}_{1}\mathcal{P}}\left(3\mathcal{D}_{1b}^{\mu}\left(\partial_{\mu}\partial_{\nu}\mathcal{P}\right)_{ba}\mathcal{D}^{*\nu\dagger}_{a}\right.\nonumber\\&-&\left.\mathcal{D}_{1b}^{\mu}\left(\partial^{\nu}\partial_{\nu}\mathcal{P}\right)_{ba}\mathcal{D}^{*\dagger}_{a\mu}\right.\nonumber\\&+&\left.\frac{1}{m_{\mathcal{D}^{*}}m_{\mathcal{D}1}}\partial^{\nu}\mathcal{D}_{1b}^{\mu}\left(\partial_{\nu}\partial_{\tau}\mathcal{P}\right)_{ba}\partial^{\tau}\mathcal{D}_{a\mu}^{*\dagger}\right),\nonumber\\
%\end{eqnarray}
%\begin{eqnarray}
\mathcal{L}_{\mathcal{D}^{(*)}\mathcal{D}_{1}\mathcal{V}}&=&ig_{\mathcal{D}^{*}\mathcal{D}_{1}\mathcal{V}}\epsilon_{\mu\nu\alpha\beta}\left(\mathcal{D}_{1b}^{\mu}\stackrel{\leftrightarrow}{\partial}^{\alpha}\mathcal{D}^{*\nu\dagger}_{a}\right)\mathcal{V}_{ba}^{\beta}\nonumber\\&+&g_{\mathcal{D}\mathcal{D}_{1}\mathcal{V}}\mathcal{D}_{b}\mathcal{D}_{1a}\mathcal{V}_{ba}^{\beta},\label{Eq:3}
\end{eqnarray}
with $\mathcal{D}^{(*)}=\left(D^{(*)0},D^{(*)+},D_{s}^{(*)+}\right)$, $\mathcal{D}_{1}=\Big(D_{1}(2420)^{0},$ $ D_{1}(2420)^{+}, D_{s1}(2536)^{+}\Big)$ and $A\stackrel{\leftrightarrow}{\partial}^\mu B=A(\partial^\mu B)-(\partial^\mu A) B $.  The $\mathcal{V}$ and $\mathcal{P}$ are the pseudoscalar and vector meson nonet in the matrices form, which are,
\begin{eqnarray}
\mathcal{V} &=&
\begin{pmatrix}
\frac{1}{\sqrt2}(\rho^{0}+\omega)&\rho^{+}&K^{*+}\\
\rho^{-}&
\frac{1}{\sqrt2}(-\rho^{0}+\omega)&K^{*0}\\
K^{*-}&\bar{K}^{*0}&\phi\\ \nonumber
\end{pmatrix},\nonumber\\
%\end{eqnarray}
%\begin{eqnarray}
\mathcal{P} &=&
\begin{pmatrix}
\frac{\pi^{0}}{\sqrt{2}}+\alpha\eta+\beta\eta^{\prime}&\pi^{+}&K^{+}\\
\pi^{-}&-\frac{\pi^{0}}{\sqrt{2}}+\alpha\eta+\beta\eta^{\prime}&K^{0}\\
K^{-}&\bar{K}^{0}&\gamma\eta+\delta\eta^{\prime}\\
\end{pmatrix}, \qquad
\end{eqnarray}
where $\alpha$, $\beta$, $\gamma$ and $\delta$ are the parameters related to the mixing angle $\theta$,  which are defined as 
\begin{eqnarray}
\alpha &=&\frac{\mathrm{cos}\theta-\sqrt{2}\mathrm{sin}\theta}{\sqrt{2}},\  \qquad \beta=\frac{\mathrm{sin}\theta+\sqrt{2}\mathrm{cos}\theta}{\sqrt{6}}\ \ \nonumber\\ 
\gamma &=&\frac{-2\mathrm{cos}\theta-\sqrt{2}\mathrm{sin}\theta}{\sqrt{6}}, \ \quad \delta=\frac{-2\mathrm{sin}\theta+\sqrt{2}\mathrm{cos}\theta}{\sqrt{6}}.\quad	
\end{eqnarray}
In the present calculations, we take the mixing angle $\theta=19.1^{\circ}$ \cite{MARK-III:1988crp,DM2:1988bfq}. %%%%

\subsection{Decay Amplitude}
With the above effective Lagrangians, we can get the amplitudes for $Y(4626)\to J/\psi\eta$, $Y(4626)\to J/\psi\eta^{\prime}$, $Y(4626)\to \eta_{c}\phi$, $Y(4626)\to \chi_{c0}\phi$, $Y(4626)\to \chi_{c1}\phi$ and $Y(4626)\to \chi_{c2}\phi$ corresponding to the diagrams in Fig.~\ref{Fig:Tri2},  which are,
\begin{eqnarray}
i\mathcal{M}_{a}&=&i^{3}\int \frac{d^{4}q}{(2\pi)^{4}}\Big[g_{Y}\varepsilon_{\mu\nu\alpha\beta}(-ip^{\mu})\epsilon^{\nu}(p)\nonumber\\&\times&\tilde{\Phi}_{Y}(-p_{12}^{2},\Lambda_{Y}^{2}) \Big]\Big[ig_{\psi D_{s}^{*}D_{s}^{*}}((iq_{\delta}+ip_{1\delta})g_{\theta\lambda} \nonumber\\
&+&(iq_{\lambda}+ip_{1\lambda})g_{\theta\delta}-(iq_{\theta}+ip_{1\theta})g_{\delta\lambda}))\epsilon^{\theta}(p_{3}) \Big] \nonumber\\
&\times&\Big[g_{D_{s}^{*}D_{s1}\eta}(3(ip_{4o})(ip_{4\xi})-g_{o\xi}(ip_{4})^{2} \nonumber\\
&+&\frac{1}{m_{D^{*}}m_{D_{1}}}(ip_{2})^{\omega}(ip_{4})_{\omega}(ip_{4})^{\upsilon}g_{o\xi}(-iq)_{\upsilon}) \Big] \nonumber\\
&\times&\frac{-g^{\beta\delta}+p_{1}^{\beta}p_{1}^{\delta}/m_{1}^{2}}{p_{1}^{2}-m_{1}^{2}}\frac{-g^{\alpha o}+p_{2}^{\alpha}p_{2}^{o}/m_{2}^{2}}{p_{2}^{2}-m_{2}^{2}}\nonumber\\
&\times&\frac{-g^{\lambda\xi}+q^{\lambda}g^{\xi}/m_{q}^{2}}{q^{2}-m_{q}^{2}} \mathcal{F}\left(m_{q}^2,\Lambda^2 \right)\nonumber,\\
i\mathcal{M}_{c}&=&i^{3}\int \frac{d^{4}q}{(2\pi)^{4}}\Big[g_{Y}\varepsilon_{\mu\nu\alpha\beta}(-ip^{\mu})\epsilon^{\nu}(p) \nonumber\\
&\times&\tilde{\Phi}_{Y}(-p_{12}^{2},\Lambda_{Y}^{2}) \Big] \Big[ig_{\eta_{c}D_{s}^{*}D_{s}^{*}}\varepsilon_{\gamma\delta\kappa\lambda}(ip_{4})^{\gamma}\nonumber\\&\times&(iq^{\kappa}+ip_{1}^{\kappa})\Big] \Big[-g_{D_{s}^{*}D_{s1}\phi}\varepsilon_{o\xi\psi\rho}(-iq^{\psi}+ip_{2}^{\psi})\nonumber\\&\times&\epsilon^{\rho}(p_{4})\Big] \frac{-g^{\beta\delta}+p_{1}^{\beta}p_{1}^{\delta}/m_{1}^{2}}{p_{1}^{2}-m_{1}^{2}}\frac{-g^{\alpha o}+p_{2}^{\alpha}p_{2}^{o}/m_{2}^{2}}{p_{2}^{2}-m_{2}^{2}}\nonumber\\&\times&\frac{-g^{\lambda\xi}+q^{\lambda}g^{\xi}/m_{q}^{2}}{q^{2}-m_{q}^{2}} \mathcal{F}\left(m_{q}^2,\Lambda^2 \right)\nonumber,\\
%\end{eqnarray}
%\begin{eqnarray}
i\mathcal{M}_{d}&=&i^{3}\int \frac{d^{4}q}{(2\pi)^{4}}\Big[g_{Y}\varepsilon_{\mu\nu\alpha\beta}(-ip^{\mu})\epsilon^{\nu}(p) \nonumber\\
&\times&\tilde{\Phi}_{Y}(-p_{12}^{2},\Lambda_{Y}^{2}) \Big] \Big[(-ig_{\eta_{c}D_{s}^{*}D_{s}}(iq_{\delta}+ip_{1\delta}) \Big] \nonumber\\
&\times&\Big[-g_{D_{s}D_{s1}\phi} \epsilon_{o}(p_{4}) \Big] \frac{-g^{\beta\delta}+p_{1}^{\beta}p_{1}^{\delta}/m_{1}^{2}}{p_{1}^{2}-m_{1}^{2}} \nonumber\\&\times&\frac{-g^{\alpha o}+p_{2}^{\alpha}p_{2}^{o}/m_{2}^{2}}{p_{2}^{2}-m_{2}^{2}}\frac{1}{q^{2}-m_{q}^{2}} \mathcal{F}\left(m_{q}^2,\Lambda^2 \right)\nonumber,%\\
\end{eqnarray} 
\begin{eqnarray}
i\mathcal{M}_{e}&=&i^{3}\int \frac{d^{4}q}{(2\pi)^{4}}\Big[g_{Y}\varepsilon_{\mu\nu\alpha\beta}(-ip^{\mu})\epsilon^{\nu}(p) \nonumber\\&\times&\tilde{\Phi}_{Y}(-p_{12}^{2},\Lambda_{Y}^{2}) \Big] \Big[g_{\chi_{c0}D_{s}^{*}D_{s}^{*}}g_{\delta\lambda} \Big] \nonumber\\&\times&\Big[-g_{D_{s}^{*}D_{s1}\phi}\varepsilon_{o\xi\upsilon\rho}(-iq^{\upsilon}+ip_{2}^{\upsilon})\epsilon^{\rho}(p_{4}) \Big] \nonumber\\&\times&\frac{-g^{\beta\delta}+p_{1}^{\beta}p_{1}^{\delta}/m_{1}^{2}}{p_{1}^{2}-m_{1}^{2}}\frac{-g^{\alpha o}+p_{2}^{\alpha}p_{2}^{o}/m_{2}^{2}}{p_{2}^{2}-m_{2}^{2}}\nonumber\\&\times &\frac{-g^{\lambda\xi}+q^{\lambda}g^{\xi}/m_{q}^{2}}{q^{2}-m_{q}^{2}} \mathcal{F}\left(m_{q}^2,\Lambda^2 \right) \nonumber,\\
%\end{eqnarray}
%\begin{eqnarray}
i\mathcal{M}_{f}&=&i^{3}\int \frac{d^{4}q}{(2\pi)^{4}}\Big[g_{Y}\varepsilon_{\mu\nu\alpha\beta}(-ip^{\mu})\epsilon^{\nu}(p) \nonumber\\
&\times&\tilde{\Phi}_{Y}(-p_{12}^{2},\Lambda_{Y}^{2}) \Big] \Big[ig_{\chi_{c1}D_{s}^{*}D_{s}}\epsilon_{\delta}(p_{3}) \Big] \nonumber\\
&\times&\Big[-g_{D_{s}D_{s1}\phi} \epsilon_{o}(p_{4}) \Big] \frac{-g^{\beta\delta}+p_{1}^{\beta}p_{1}^{\delta}/m_{1}^{2}}{p_{1}^{2}-m_{1}^{2}} \nonumber\\
&\times& \frac{-g^{\alpha o}+p_{2}^{\alpha}p_{2}^{0}/m_{2}^{2}}{p_{2}^{2}-m_{2}^{2}} \frac{1}{q^{2}-m_{q}^{2}} \mathcal{F}\left(m_{q}^2,\Lambda^2 \right),\nonumber\\
%\end{eqnarray}
%\begin{eqnarray}
i\mathcal{M}_{g}&=&i^{3}\int \frac{d^{4}q}{(2\pi)^{4}}\Big[g_{Y}\varepsilon_{\mu\nu\alpha\beta}(-ip^{\mu})\epsilon^{\nu}(p) \nonumber\\
&\times&\tilde{\Phi}_{Y}(-p_{12}^{2},\Lambda_{Y}^{2}) \Big] \Big[g_{\chi_{c2}D_{s}^{*}D_{s}^{*}}\epsilon_{\delta\lambda}(p_{3}) \Big] \nonumber\\
&\times&\Big[-g_{D_{s}^{*}D_{s1}\phi}\varepsilon_{o\xi\upsilon\rho}(-iq^{\upsilon}+ip_{2}^{\upsilon})\epsilon^{\rho}(p_{4}) \Big] \nonumber\\&\times&\frac{-g^{\beta\delta}+p_{1}^{\beta}p_{1}^{\delta}/m_{1}^{2}}{p_{1}^{2}-m_{1}^{2}}\frac{-g^{\alpha o}+p_{2}^{\alpha}p_{2}^{o}/m_{2}^{2}}{p_{2}^{2}-m_{2}^{2}}\nonumber\\&&\times\frac{-g^{\lambda\xi}+q^{\lambda}g^{\xi}/m_{q}^{2}}{q^{2}-m_{q}^{2}} \mathcal{F}\left(m_{q}^2,\Lambda^2 \right).
\end{eqnarray}

In the above amplitudes, we introduce a monopole form factor not only to describe the exchanging mesons inner structure but also avoid the divergence in the loop integrals at the ultraviolet region\cite{Tornqvist:1993ng,Wu:2021udi,Wu:2021caw,Wu:2023fyh,Cheng:2004ru,Gortchakov:1995im}, which is,
\begin{eqnarray}
\mathcal{F}\left(m_{q}^2,\Lambda^2\right)=\frac{m_{q}^{2}-\Lambda^{2}}{q^{2}-\Lambda^{2}},
\end{eqnarray}
where the parameter $\Lambda$ can be further parameterized to be $\Lambda=m_{q}+\alpha\Lambda_{QCD}$ with $\Lambda_{QCD}=0.22~\mathrm{GeV}$ and $m_{q}$ is the mass of the exchanging mesons. The parameter $\alpha$ is typically of the order of unity.

The amplitude for $\mathcal{M}_{b}$ can be obtained by replacing $m_{\eta}$, $g_{D_{s}^{*}D_{s1}\eta}$ in $\mathcal{M}_{a}$ with $m_{\eta^{\prime}}$, $g_{D_{s}^{*}D_{s1}\eta^{\prime}}$, and the total amplitudes for each channel are,
\begin{eqnarray}
\mathcal{M}_{Y\to J/\psi\eta}&=&2 \mathcal{M}_{a},\nonumber\\
\mathcal{M}_{Y\to J/\psi\eta^{\prime}}&=&2 \mathcal{M}_{b},\nonumber\\
\mathcal{M}_{Y\to \eta_{c}\phi}&=& 2(\mathcal{M}_{c}+\mathcal{M}_{d}),\nonumber\\
\mathcal{M}_{Y\to \chi_{c0}\phi}&=&2\mathcal{M}_{e},\nonumber\\
\mathcal{M}_{Y\to \chi_{c1}\phi}&=&2\mathcal{M}_{f},\nonumber\\
\mathcal{M}_{Y\to \chi_{c2}\phi}&=&2\mathcal{M}_{g},
\end{eqnarray}
where the factor 2 comes from charge symmetry.

With the amplitudes discussed above, the partial width of the decay processes could be calculated by,
\begin{eqnarray}
\Gamma_{Y\to...}=\frac{1}{3}\frac{1}{8\pi}\frac{|\vec{p}|}{m_{X}^{2}}\left|\ \overline{\mathcal{M}_{X\to...}}\ \right|^{2}.\label{Eq:PW}
\end{eqnarray}
where the factor $1/3$ is resulted from the average of the spin of $Y(4626)$, and the overline indicates the sum over the spin of the involved particles.

\section{Numerical results and discussions}
\label{sec:Sec4}

\begin{figure}[t]
 \centering
\includegraphics[width=8.2cm]{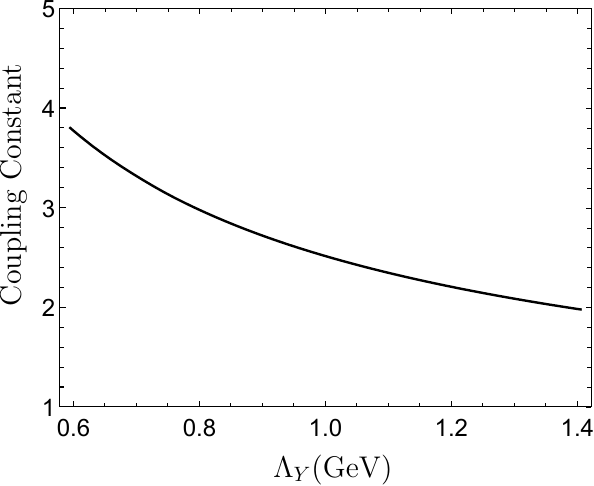}
  \caption{The Coupling constant $g_{Y}$ depending on the model parameter $\Lambda_{Y}$.}\label{Fig:Tri3}
\end{figure}

\label{sec:Sec4}
\subsection{Coupling Constants}
Before we estimate the hidden charm decays of  $Y(4626)$, the relevant coupling constants should be clarified further. The coupling constants of the charmonia and $S$-wave  charmed/charmed-strange mesons are well determined in the heavy quark limit, which are~\cite{Casalbuoni:1996pg,Cheng:1992xi,Yan:1992gz,Wise:1992hn,Wu:2021udi}, 
\begin{eqnarray}
g_{\psi \mathcal{D}\mathcal{D}}&=&2g_{1}\sqrt{m_{\psi}}m_{\mathcal{D}}\nonumber\\
g_{\psi \mathcal{D}^{*}\mathcal{D}}&=&2g_{1}\sqrt{{m_{\mathcal{D}^{*}}{m_{\mathcal{D}}}}/m_{\psi}}\nonumber\\
g_{\psi \mathcal{D}^{*}\mathcal{D}^{*}}&=&2g_{1}\sqrt{m_{\psi}}m_{\mathcal{D}^{*}}\nonumber\\
g_{\eta_{c}\mathcal{D}^{*}\mathcal{D}}&=&2g_{1}\sqrt{m_{\mathcal{D}}m_{\mathcal{D}^{*}}m_{\eta}}\nonumber\\
g_{\eta_{c}\mathcal{D}^{*}\mathcal{D}^{*}}&=&2g_{1}m_{\mathcal{D}^{*}}/\sqrt{m_{\eta_{c}}}\nonumber\\
g_{\chi_{c0}\mathcal{D}\mathcal{D}}&=&-2\sqrt{3}g_{2}\sqrt{m_{\chi_{c0}}}m_{\mathcal{D}}\nonumber\\
g_{\chi_{c0}\mathcal{D}^{*}\mathcal{D}^{*}}&=&-2g_{2}\sqrt{m_{\chi_{c0}}}m_{\mathcal{D}^{*}}/\sqrt{3}\nonumber\\
g_{\chi_{c1}\mathcal{D}^{*}\mathcal{D}}&=&2\sqrt{2}g_{2}\sqrt{2m_{\chi_{c1}}{m_{\mathcal{D}}}{m_{\mathcal{D}^{*}}}}\nonumber\\
g_{\chi_{c2}\mathcal{D}^{*}\mathcal{D}^{*}}&=&4g_{2}\sqrt{m_{\chi_{c2}}}m_{\mathcal{D}^{*}}
\end{eqnarray}
with $g_{1}=\sqrt{m_{\psi}}/({2m_{D}f_{\psi}})$, and $f_{\psi}=405~\mathrm{MeV}$ to be the decay constant of $J/\psi$, while $g_{2}=-\sqrt{m_{\chi_{c0}}/3}/f_{\chi_{c0}}$, $f_{\chi_{c0}}=510~\mathrm{MeV}$ to be decay constant of $\chi_{c0}$ \cite{Colangelo:2002mj}.

For the couplings constants related to the charmed (charmed-strange) mesons and the light mesons, they are determined by heavy quark limit and chiral symmetry, and the relevant coupling constants read~ \cite{He:2019csk,Casalbuoni:1996pg,Wu:2021udi},%%%
\begin{eqnarray}
g_{\mathcal{D}^{*}\mathcal{D}_{1}\mathcal{P}}&=&-\frac{\sqrt{6}}{3}\frac{h_{1}+h_{2}}{\Lambda_{\chi}f_{\pi}}\sqrt{m_{\mathcal{D}_{1}}m_{\mathcal{D}^{*}}}\nonumber\\
g_{\mathcal{D} \mathcal{D}_1 \mathcal{V}} &=&\frac{2g_{V}\xi_{1}}{\sqrt{3}} \sqrt{m_{\mathcal{D}} m_{\mathcal{D}_1}},\nonumber\\
g_{\mathcal{D}^\ast \mathcal{D}_1 \mathcal{V}} &=&\frac{g_{V}\xi_{1}}{\sqrt{3}},
\end{eqnarray}
with $(h_{1}+h_{2})/\Lambda_{\chi}=0.55 \ \mathrm{GeV}^{-1} $, $ 
\xi_{1}=-0.1$ and $g_{V}=5.9$.  

Finally, the coupling constant of $Y(4626)$ with its constituents, $g_{Y}$, could be estimated using the compositeness condition provided in Eq.~(\ref{Eq:2}). Here, $\Lambda_{Y}$ is a phenomenological parameter, which should be of the order of $1~\mathrm{GeV}$. In the present estimations, we vary the parameter $\Lambda_{Y}$ from $0.6~\mathrm{GeV}$ to $1.4~\mathrm{GeV}$ to check the model parameter dependences of the coupling constant and the partial widths. The $ \Lambda_{Y}$ dependence of the coupling constant $g_{Y}$ is presented in Fig.~\ref{Fig:Tri3}. From the figure, one can find that the coupling constant monotonically decreases with increasing model parameter $\Lambda_{Y}$. In particular, the coupling decreases from $3.78$ to $1.98$ in the considered parameter range.

\subsection{Partial Widths Of The Hidden Charm Decays}

\begin{figure*}[htb]
\begin{tabular}{cc}
\centering
\includegraphics[width=8.2cm]{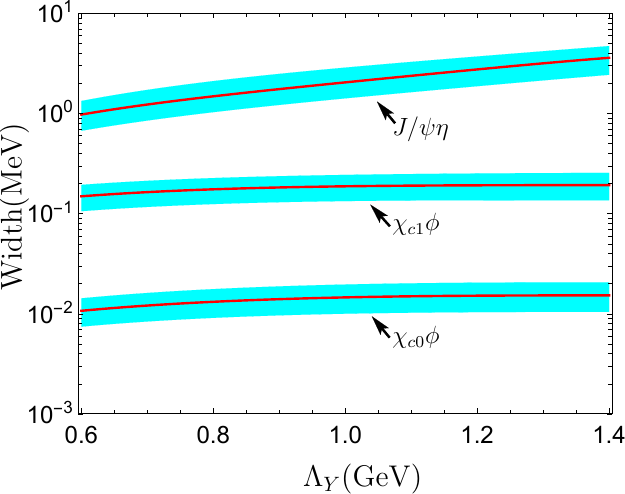}~~~~
\includegraphics[width=8.2cm]{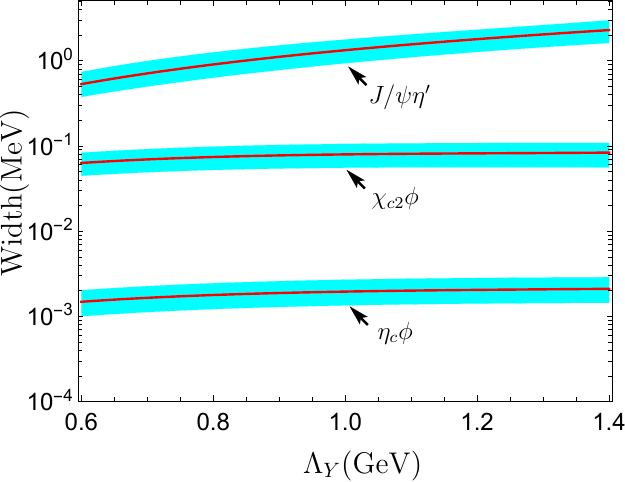}
\end{tabular}
\caption{The partial widths of $Y(4626)\to J/\psi\eta$,~$J/\psi\eta^{\prime}$,~$\eta_{c}\phi$,~$\chi_{c0}\phi$,~$\chi_{c1}\phi$ and $\chi_{c2}\phi$ depending on the model parameters $\Lambda_{Y}$. The red solid curves are obtained with $\alpha=1.0$, while the cyan bands indicate the uncertainties resulted from $\alpha$ variation from $0.8$ to $1.2$.}\label{Fig:Tri4}
\end{figure*}

\renewcommand{\arraystretch}{1.5} 
\begin{table}
\caption{The partial widths of the hidden charm decays of $Y(4626)$. The uncertainties are resulted from variations of the model parameters $\Lambda_Y$ and $\alpha$. \label{Tab:PW}}
\begin{tabular}{p{1.5cm}<\centering p{2.5cm}<\centering |p{1.5cm}<\centering p{2.5cm}<\centering }
\toprule[1pt]
Process & Width (MeV) & Process & Width (MeV) \\
\midrule[1pt]
$Y\to J/\psi \eta$  &  $0.97\sim 3.58$ &
$Y\to J/\psi \eta^\prime $ & $0.53\sim 2.27$\\
 $Y\to \eta_{c}\phi$ & $(1.48-2.09)\times 10^{-3}$ &
$Y\to \chi_{c0}\phi$ & $(1.07\sim 1.52)\times 10^{-2}$ \\
$Y\to \chi_{c1}\phi$ & $0.15\sim 0.19$ &
$Y\to \chi_{c2}\phi$ & $(6.26-8.29)\times 10^{-2}$\\
\bottomrule[1pt]
\end{tabular}
\end{table}

With the above preparations, we could estimate the partial width using Eq.~\eqref{Eq:PW}. In the present work, there are two model parameters, which are $\Lambda_Y$ and $\alpha$ introduced by the correlation functions of the molecule and by the form factors, respectively. For $\Lambda_Y$, it varies from $0.6~\mathrm{GeV}$ to $1.4~\mathrm{GeV}$, while for $\alpha$, we take three representative values, which are $0.8$, $1.0$, and $1.2$, respectively. 

In Fig.~\ref{Fig:Tri4}, we illustrate the partial widths of the considered hidden charm decay processes depending on the model parameters $\Lambda_{Y}$ and $\alpha$. The red curves are the partial widths  estimated with $\alpha=1.0$, while the cyan bands indicate the uncertainties resulting from the variation of $\alpha$ from $0.8$ to $1.2$. From the figure, one can find that the partial widths increase smoothly with increasing $\Lambda_Y$. In particular, the partial width of $Y(4626) \to J/\psi \eta $ is the largest among the considered hidden charm decay processes, which is estimated to be $(0.97\sim 3.58)$ MeV in the considered parameters range, indicating that the branching fraction of $Y(4626)\to J/\psi \eta $ should be several percent. Another process with a large partial width is $Y(4626)\to J/\psi \eta^\prime$ with a partial width of $(0.53 \sim 2.27)$ MeV, which indicate that branching ratio of this process should be greater than $1\%$. For the process $Y(4260)\to \chi_{c1} \phi$, the partial width is estimated to be of the order of 0.1 MeV with the corresponding branching fraction of $10^{-3}$. The widths of the processes $Y(4626)\to \chi_{c0}\phi$, $\chi_{c2}\phi$ and $\eta_c \phi$ are even smaller, which are of the order $10^{-3}$ MeV or $10^{-2}$ MeV. The concrete values of the estimated partial widths of the considered hidden charm decay processes are collected in Table \ref{Tab:PW}.

\begin{figure}[t]
\centering
\includegraphics[width=8.5cm]{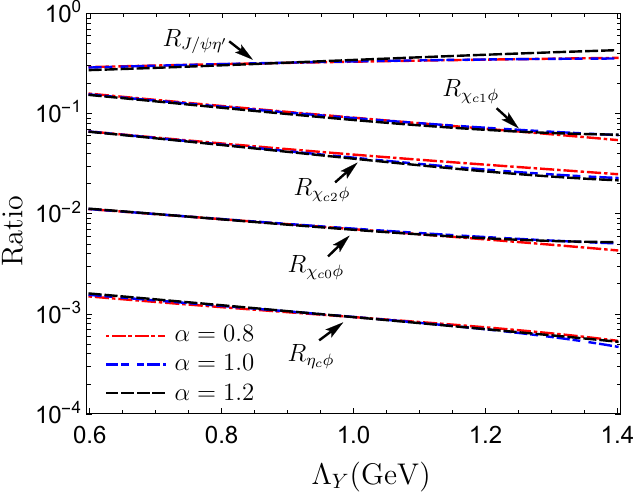}
\caption{The ratio $R_{AB}$ depending on the model parameters $\Lambda$ with typical values of $\alpha$.}
\label{Fig:Tri5}
\end{figure}

Moreover, from Fig.~\ref{Fig:Tri4}, one can find that the model parameter dependences of the estimated partial widths for different hidden charm decay processes are very similar, and thus their ratios are expected to be weakly dependent on the model parameters. Here, we define the ratio $R_{AB}$ as,
\begin{eqnarray}
	R_{AB}= \frac{\Gamma_{Y(4260)\to AB}}{\Gamma_{Y(4260)\to J/\psi \eta}}.
\end{eqnarray}
In Fig.~\ref{Fig:Tri5}, we present the decay widths ratios depending on the model parameter $\Lambda_Y$ with three typical values of $\alpha$, which are $0.8$, $1.0$ and $1.2$, respectively. From the figure one can find that the curves for a certain ratio with different $\alpha$ are almost degenerated, which indicates that the ratios are almost independent on the model parameter $\alpha$. As for the $\Lambda_Y$ dependences, we find $R_{J/\psi \eta^\prime}$ increases from 0.54 to 0.63 with $\Lambda_Y$ increasing from 0.6 to 1.4, while $R_{\chi_{c1}\phi}$ decreases with $\Lambda_Y$ increasing, and in the considered $\Lambda_Y$ range, $R_{\chi_{c1}\phi}$ is estimated to be $0.05 \sim 0.16$. The other ratios, $R_{\chi_{c0}\phi}$, $R_{\chi_{c2} \phi}$ and $R_{\eta_c \phi}$, are estimated to be less than $10^{-1}$. 

It is worth mentioning that the cross sections for $e^+ e^- \to J/\psi \eta$ have been measured by the BES III Collaboration~\cite{BESIII:2023tll}. However, the cross sections above 4.46 GeV were not precisely measured, and only the upper limits were reported. One can find that in the vicinity of $Y(4626)$ the upper limits of the cross section were measured to be $0.01$, $1.78$, $1.92$, and $0.84$ pb for $\sqrt{s}=4.5995$, $4.6119$, $4.6280$, and $4.6409$ GeV, respectively. To further investigate $Y(4626)$ in the cross sections for $e^+ e^- \to \eta J/\psi$, more precise measurements are needed, which should be accessible by the BES III and Belle II Collaborations.

\section{Summary}
\label{sec:Sec5}
Stimulated by the observation of the charmonium-like state $Y(4626)$ in the process $e^{+}e^{-}\to D_{s}^{+}D_{s1}(2536)^{-}+c.c.$ by Belle Collaboration, and the fact that the observed mass of $Y(4626)$ is very close to the threshold of $D_{s}^{*+}D_{s1}(2536)^{-}$, the $D_s^{\ast+} D_{s1}(2536)^-$ molecular interpretation has been proposed in the literature. 

In the present work, we investigate the hidden charm decay processes of $Y(4626)$ in the molecular frame by the effective Lagrangian approach. The partial widths of $Y(4626)\to J/\psi\eta$, $J/\psi\eta^{\prime}$, $\eta_{c}\phi$, $\chi_{c0}\phi$, $\chi_{c1}\phi$ and $\chi_{c2}\phi$ are estimated. Our estimation indicates that the partial width for the $J/\psi\eta$ and $J/\psi\eta^{\prime}$ channels can be of the order of 1 MeV, while the one for $Y(4626)\to \chi_{c1} \phi$ is of the order of 0.1 MeV. Based on the present estimations, we propose to search $Y(4626)$ in the $e^+ e^- \to J/\psi \eta$ and $e^+e^-\to J/\psi \eta^\prime$ processes, which should be accessible by BES \uppercase\expandafter{\romannumeral 3} and Belle \uppercase\expandafter{\romannumeral 2}.

\section{ACKNOWLEDGMENTS} 
This work is supported by the National Natural Science Foundation of China under Grants Nos. 12175037 and 12335001. Zi-Li Yue is also supported by the SEU Innovation Capability Enhancement Plan for Doctoral Students (Grants no. CXJH$\_$SEU 24135).


\begin{thebibliography}{00}
\bibitem{BESIII:2013ris}
M.~Ablikim \textit{et al.} [BESIII],
%``Observation of a Charged Charmoniumlike Structure in $e^+e^- \to \pi^+\pi^- J/\psi$ at $\sqrt{s}$ =4.26 GeV,''
Phys. Rev. Lett. \textbf{110} (2013), 252001
doi:10.1103/PhysRevLett.110.252001
[arXiv:1303.5949 [hep-ex]].
%1003 citations counted in INSPIRE as of 05 Feb 2023
\bibitem{BESIII:2013ouc}
M.~Ablikim \textit{et al.} [BESIII],
%``Observation of a Charged Charmoniumlike Structure $Z_c$(4020) and Search for the $Z_c$(3900) in $e^+e^- \to \pi^+\pi^-h_c$,''
Phys. Rev. Lett. \textbf{111} (2013) no.24, 242001
doi:10.1103/PhysRevLett.111.242001
[arXiv:1309.1896 [hep-ex]].
%474 citations counted in INSPIRE as of 05 Feb 2023
\bibitem{CDF:2009jgo}
T.~Aaltonen \textit{et al.} [CDF],
%``Evidence for a Narrow Near-Threshold Structure in the $J/\psi\phi$ Mass Spectrum in $B^+\to J/\psi\phi K^+$ Decays,''
Phys. Rev. Lett. \textbf{102} (2009), 242002
doi:10.1103/PhysRevLett.102.242002
[arXiv:0903.2229 [hep-ex]].
%402 citations counted in INSPIRE as of 05 Feb 2023
\bibitem{BaBar:2006ait}
B.~Aubert \textit{et al.} [BaBar],
%``Evidence of a broad structure at an invariant mass of 4.32- $GeV/c^{2}$ in the reaction $e^{+} e^{-} \to \pi^{+} \pi^{-} \psi_{2S}$ measured at BaBar,''
Phys. Rev. Lett. \textbf{98} (2007), 212001
doi:10.1103/PhysRevLett.98.212001
[arXiv:hep-ex/0610057 [hep-ex]].
%395 citations counted in INSPIRE as of 05 Feb 2023
\bibitem{Belle:2007umv}
X.~L.~Wang \textit{et al.} [Belle],
%``Observation of Two Resonant Structures in e+e- to pi+ pi- psi(2S) via Initial State Radiation at Belle,''
Phys. Rev. Lett. \textbf{99} (2007), 142002
doi:10.1103/PhysRevLett.99.142002
[arXiv:0707.3699 [hep-ex]].
%498 citations counted in INSPIRE as of 05 Feb 2023
\bibitem{Belle:2014nuw}
K.~Chilikin \textit{et al.} [Belle],
%``Observation of a new charged charmoniumlike state in $\bar{B}^0 \to J/\psi K^- \pi^+$ decays,''
Phys. Rev. D \textbf{90} (2014) no.11, 112009
doi:10.1103/PhysRevD.90.112009
[arXiv:1408.6457 [hep-ex]].
%205 citations counted in INSPIRE as of 05 Feb 2023
\bibitem{D0:2013jvp}
V.~M.~Abazov \textit{et al.} [D0],
%``Search for the $X$(4140) state in $B^+ \to $J$_{\psi,\phi}K^+$ decays with the D0 Detector,''
Phys. Rev. D \textbf{89} (2014) no.1, 012004
doi:10.1103/PhysRevD.89.012004
[arXiv:1309.6580 [hep-ex]].
%134 citations counted in INSPIRE as of 14 Nov 2023
\bibitem{CMS:2013jru}
S.~Chatrchyan \textit{et al.} [CMS],
%``Observation of a Peaking Structure in the $J/\psi \phi$ Mass Spectrum from $B^{\pm} \to J/\psi \phi K^{\pm}$ Decays,''
Phys. Lett. B \textbf{734} (2014), 261-281
doi:10.1016/j.physletb.2014.05.055
[arXiv:1309.6920 [hep-ex]].
%190 citations counted in INSPIRE as of 14 Nov 2023
\bibitem{LHCb:2016axx}
R.~Aaij \textit{et al.} [LHCb],
%``Observation of $J/\psi\phi$ structures consistent with exotic states from amplitude analysis of $B^+\to J/\psi \phi K^+$ decays,''
Phys. Rev. Lett. \textbf{118} (2017) no.2, 022003
doi:10.1103/PhysRevLett.118.022003
[arXiv:1606.07895 [hep-ex]].
%291 citations counted in INSPIRE as of 14 Nov 2023
\bibitem{LHCb:2021uow}
R.~Aaij \textit{et al.} [LHCb],
%``Observation of New Resonances Decaying to $J/\psi K^+$+ and $J/\psi \phi$,''
Phys. Rev. Lett. \textbf{127} (2021) no.8, 082001
doi:10.1103/PhysRevLett.127.082001
[arXiv:2103.01803 [hep-ex]].
%169 citations counted in INSPIRE as of 14 Nov 2023


%============================================JISUAN================================================================================%%%%%
\bibitem{BaBar:2012hpr}
J.~P.~Lees \textit{et al.} [BaBar],
%``Study of the reaction $e^{+}e^{-}\to \psi(2S)\pi^{-}\pi^{-}$ via initial-state radiation at BaBar,''
Phys. Rev. D \textbf{89} (2014) no.11, 111103
doi:10.1103/PhysRevD.89.111103
[arXiv:1211.6271 [hep-ex]].
%142 citations counted in INSPIRE as of 14 Nov 2023
\bibitem{Belle:2003nnu}
S.~K.~Choi \textit{et al.} [Belle],
%``Observation of a narrow charmonium-like state in exclusive $B^\pm \to K^\pm \pi^+ \pi^- J/\psi$ decays,''
Phys. Rev. Lett. \textbf{91} (2003), 262001
doi:10.1103/PhysRevLett.91.262001
[arXiv:hep-ex/0309032 [hep-ex]].
%2261 citations counted in INSPIRE as of 05 Feb 2023
\bibitem{BaBar:2005hhc}
B.~Aubert \textit{et al.} [BaBar],
%``Observation of a broad structure in the $\pi^+ \pi^- J/\psi$ mass spectrum around 4.26-GeV/c$^2$,''
Phys. Rev. Lett. \textbf{95} (2005), 142001
doi:10.1103/PhysRevLett.95.142001
[arXiv:hep-ex/0506081 [hep-ex]].
%1006 citations counted in INSPIRE as of 14 Nov 2023
\bibitem{CLEO:2006tct}
Q.~He \textit{et al.} [CLEO],
%``Confirmation of the Y(4260) resonance production in ISR,''
Phys. Rev. D \textbf{74} (2006), 091104
doi:10.1103/PhysRevD.74.091104
[arXiv:hep-ex/0611021 [hep-ex]].
%285 citations counted in INSPIRE as of 21 Apr 2024
\bibitem{Belle:2013yex}
Z.~Q.~Liu \textit{et al.} [Belle],
%``Study of $e^+e^- \to \pi^+ \pi^- J/\psi $ and Observation of a Charged Charmoniumlike State at Belle,''
Phys. Rev. Lett. \textbf{110} (2013), 252002
[erratum: Phys. Rev. Lett. \textbf{111} (2013), 019901]
doi:10.1103/PhysRevLett.110.252002
[arXiv:1304.0121 [hep-ex]].
%889 citations counted in INSPIRE as of 21 Apr 2024
\bibitem{Belle:2007dxy}
C.~Z.~Yuan \textit{et al.} [Belle],
%``Measurement of e+ e- ---\ensuremath{>} pi+ pi- J/psi cross-section via initial state radiation at Belle,''
Phys. Rev. Lett. \textbf{99} (2007), 182004
doi:10.1103/PhysRevLett.99.182004
[arXiv:0707.2541 [hep-ex]].
%526 citations counted in INSPIRE as of 21 Apr 2024
\bibitem{BaBar:2012vyb}
J.~P.~Lees \textit{et al.} [BaBar],
%``Study of the reaction $e^{+}e^{-} \to J/\psi\pi^{+}\pi^{-}$ via initial-state radiation at BaBar,''
Phys. Rev. D \textbf{86} (2012), 051102
doi:10.1103/PhysRevD.86.051102
[arXiv:1204.2158 [hep-ex]].
%152 citations counted in INSPIRE as of 21 Apr 2024

%\cite{BESIII:2016bnd}
\bibitem{BESIII:2020oph}
M.~Ablikim \textit{et al.} [BESIII],
%``Study of the process $e^+e^-\to\pi^0\pi^0 J/\psi$ and neutral charmonium-like state $Z_c(3900)^0$,''
Phys. Rev. D \textbf{102} (2020) no.1, 012009
doi:10.1103/PhysRevD.102.012009
[arXiv:2004.13788 [hep-ex]].
%43 citations counted in INSPIRE as of 01 Aug 2024


\bibitem{BESIII:2016bnd}
M.~Ablikim \textit{et al.} [BESIII],
%``Precise measurement of the $e^+e^-\to \pi^+\pi^-J/\psi$ cross section at center-of-mass energies from 3.77 to 4.60 GeV,''
Phys. Rev. Lett. \textbf{118} (2017) no.9, 092001
doi:10.1103/PhysRevLett.118.092001
[arXiv:1611.01317 [hep-ex]].
%270 citations counted in INSPIRE as of 01 Aug 2024

%\cite{BESIII:2020oph}
\bibitem{ParticleDataGroup:2022pth}
R.~L.~Workman \textit{et al.} [Particle Data Group],
%``Review of Particle Physics,''
PTEP \textbf{2022} (2022), 083C01
doi:10.1093/ptep/ptac097
%2896 citations counted in INSPIRE as of 21 Apr 2024

%\cite{Chen:2010nv}
\bibitem{Chen:2010nv}
D.~Y.~Chen, J.~He and X.~Liu,
%``Nonresonant explanation for the Y(4260) structure observed in the $e^+e^-\to J/\psi\pi^+\pi^-$ process,''
Phys. Rev. D \textbf{83} (2011), 054021
doi:10.1103/PhysRevD.83.054021
[arXiv:1012.5362 [hep-ph]].
%41 citations counted in INSPIRE as of 08 Aug 2024


\bibitem{Chen:2015bft}
D.~Y.~Chen, X.~Liu, X.~Q.~Li and H.~W.~Ke,
%``Unified Fano-like interference picture for charmoniumlike states Y(4008), Y(4260) and Y(4360),''
Phys. Rev. D \textbf{93} (2016), 014011
doi:10.1103/PhysRevD.93.014011
[arXiv:1512.04157 [hep-ph]].
%23 citations counted in INSPIRE as of 21 Apr 2024


%\cite{Chen:2016byt}
\bibitem{Chen:2016byt}
D.~Y.~Chen, Y.~B.~Dong, M.~T.~Li and W.~L.~Wang,
%``Pionic transition from Y(4260) to Z$_{c}$(3900) in a hadronic molecular scenario,''
Eur. Phys. J. A \textbf{52} (2016) no.10, 310
doi:10.1140/epja/i2016-16310-0
%12 citations counted in INSPIRE as of 08 Aug 2024

\bibitem{Li:2013bca}
M.~T.~Li, W.~L.~Wang, Y.~B.~Dong and Z.~Y.~Zhang,
%``A Study of P-wave Heavy Meson Interactions in A Chiral Quark Model,''
[arXiv:1303.4140 [nucl-th]].
%25 citations counted in INSPIRE as of 21 Apr 2024
\bibitem{Chiu:2005ey}
T.~W.~Chiu \textit{et al.} [TWQCD],
%``Y(4260) on the lattice,''
Phys. Rev. D \textbf{73} (2006), 094510
doi:10.1103/PhysRevD.73.094510
[arXiv:hep-lat/0512029 [hep-lat]].
%60 citations counted in INSPIRE as of 21 Apr 2024
\bibitem{Ding:2008gr}
G.~J.~Ding,
%``Are Y(4260) and Z+(2) are D(1) D or D(0) D* Hadronic Molecules?,''
Phys. Rev. D \textbf{79} (2009), 014001
doi:10.1103/PhysRevD.79.014001
[arXiv:0809.4818 [hep-ph]].
%190 citations counted in INSPIRE as of 14 Nov 2023


\bibitem{Close:2009ag}
F.~Close and C.~Downum,
%``On the possibility of Deeply Bound Hadronic Molecules from single Pion Exchange,''
Phys. Rev. Lett. \textbf{102} (2009), 242003
doi:10.1103/PhysRevLett.102.242003
[arXiv:0905.2687 [hep-ph]].
%55 citations counted in INSPIRE as of 06 Aug 2024
\bibitem{Close:2010wq}
F.~Close, C.~Downum and C.~E.~Thomas,
%``Novel Charmonium and Bottomonium Spectroscopies due to Deeply Bound Hadronic Molecules from Single Pion Exchange,''
Phys. Rev. D \textbf{81} (2010), 074033
doi:10.1103/PhysRevD.81.074033
[arXiv:1001.2553 [hep-ph]].
%63 citations counted in INSPIRE as of 06 Aug 2024


%\cite{Chen:2013bha}
\bibitem{Dong:2019ofp}
X.~K.~Dong, Y.~H.~Lin and B.~S.~Zou,
%``Prediction of an exotic state around 4240 MeV with $J^{PC}=1^{-+}$ as C-parity partner of Y(4260) in molecular picture,''
Phys. Rev. D \textbf{101} (2020) no.7, 076003
doi:10.1103/PhysRevD.101.076003
[arXiv:1910.14455 [hep-ph]].
%28 citations counted in INSPIRE as of 06 Aug 2024
\bibitem{Wang:2020lua}
Z.~Y.~Wang, J.~J.~Qi, J.~Xu and X.~H.~Guo,
%``Studying the $D_1D$ molecule in the Bethe-Salpeter equation approach,''
Phys. Rev. D \textbf{102} (2020) no.3, 036008
doi:10.1103/PhysRevD.102.036008
[arXiv:2004.14085 [hep-ph]].
%7 citations counted in INSPIRE as of 06 Aug 2024
\bibitem{Chen:2013bha}
D.~Y.~Chen, X.~Liu and T.~Matsuki,
%``Novel charged charmoniumlike structures in the hidden-charm dipion decays of Y(4360),''
Phys. Rev. D \textbf{88} (2013) no.1, 014034
doi:10.1103/PhysRevD.88.014034
[arXiv:1306.2080 [hep-ph]].
%25 citations counted in INSPIRE as of 08 Aug 2024


\bibitem{Wang:2021ajy}
F.~L.~Wang, R.~Chen and X.~Liu,
%``A new group of doubly charmed molecule with T-doublet charmed meson pair,''
Phys. Lett. B \textbf{835} (2022), 137502
doi:10.1016/j.physletb.2022.137502
[arXiv:2111.00208 [hep-ph]].
%19 citations counted in INSPIRE as of 06 Aug 2024
\bibitem{Belle:2019qoi}
S.~Jia \textit{et al.} [Belle],
%``Observation of a vector charmoniumlike state in $e^+e^- \to D^+_sD_{s1}(2536)^-+c.c.$,''
Phys. Rev. D \textbf{100} (2019) no.11, 111103
doi:10.1103/PhysRevD.100.111103
[arXiv:1911.00671 [hep-ex]].
%28 citations counted in INSPIRE as of 05 Feb 2023
\bibitem{Belle:2020wtd}
S.~Jia \textit{et al.} [Belle],
%``Evidence for a vector charmoniumlike state in $e^+e^- \to D^+_sD^*_{s2}(2573)^-+c.c.$,''
Phys. Rev. D \textbf{101} (2020) no.9, 091101
doi:10.1103/PhysRevD.101.091101
[arXiv:2004.02404 [hep-ex]].
%20 citations counted in INSPIRE as of 13 Aug 2024
\bibitem{Belle:2008xmh}
G.~Pakhlova \textit{et al.} [Belle],
%``Observation of a near-threshold enhancement in the e+e- ---\ensuremath{>} Lambda+(c) Lambda-(c) cross section using initial-state radiation,''
Phys. Rev. Lett. \textbf{101} (2008), 172001
doi:10.1103/PhysRevLett.101.172001
[arXiv:0807.4458 [hep-ex]].
%287 citations counted in INSPIRE as of 14 Nov 2023
\bibitem{Bugg:2008sk}
D.~V.~Bugg,
%``An Alternative fit to Belle mass spectra for D anti-D, D* anti-D* and Lambda(C) anti-Lambda(c),''
J. Phys. G \textbf{36} (2009), 075002
doi:10.1088/0954-3899/36/7/075002
[arXiv:0811.2559 [hep-ph]].
%31 citations counted in INSPIRE as of 14 Nov 2023
\bibitem{Guo:2010tk}
F.~K.~Guo, J.~Haidenbauer, C.~Hanhart and U.~G.~Meissner,
%``Reconciling the X(4630) with the Y(4660),''
Phys. Rev. D \textbf{82} (2010), 094008
doi:10.1103/PhysRevD.82.094008
[arXiv:1005.2055 [hep-ph]].
%61 citations counted in INSPIRE as of 14 Nov 2023
\bibitem{Song:2022yfr}
L.~Q.~Song, D.~Song, J.~T.~Zhu and J.~He,
%``Possible \ensuremath{\Lambda}c\ensuremath{\Lambda}\textasciimacron{}c molecular states and their productions in nucleon-antinucleon collision,''
Phys. Lett. B \textbf{835} (2022), 137586
doi:10.1016/j.physletb.2022.137586
[arXiv:2207.13957 [hep-ph]].
%5 citations counted in INSPIRE as of 14 Nov 2023
\bibitem{Cotugno:2009ys}
G.~Cotugno, R.~Faccini, A.~D.~Polosa and C.~Sabelli,
%``Charmed Baryonium,''
Phys. Rev. Lett. \textbf{104} (2010), 132005
doi:10.1103/PhysRevLett.104.132005
[arXiv:0911.2178 [hep-ph]].
%106 citations counted in INSPIRE as of 14 Nov 2023
\bibitem{Zhang:2020gtx}
J.~R.~Zhang,
%``$Y(4626)$ as a $P$-wave $[cs][\bar{c}\bar{s}]$ tetraquark state,''
Phys. Rev. D \textbf{102} (2020) no.5, 054006
doi:10.1103/PhysRevD.102.054006
[arXiv:2004.10985 [hep-ph]].
%3 citations counted in INSPIRE as of 14 Nov 2023
\bibitem{Deng:2019dbg}
C.~Deng, H.~Chen and J.~Ping,
%``Can the state $Y(4626)$ be a $P$-wave tetraquark state $[cs][\bar{c}\bar{s}]$?,''
Phys. Rev. D \textbf{101} (2020) no.5, 054039
doi:10.1103/PhysRevD.101.054039
[arXiv:1912.07174 [hep-ph]].
%18 citations counted in INSPIRE as of 14 Nov 2023
\bibitem{Ding:2007rg}
G.~J.~Ding, J.~J.~Zhu and M.~L.~Yan,
%``Canonical Charmonium Interpretation for Y(4360) and Y(4660),''
Phys. Rev. D \textbf{77} (2008), 014033
doi:10.1103/PhysRevD.77.014033
[arXiv:0708.3712 [hep-ph]].
%81 citations counted in INSPIRE as of 14 Nov 2023
\bibitem{Wang:2020prx}
J.~Z.~Wang, R.~Q.~Qian, X.~Liu and T.~Matsuki,
%``Are the $Y$ states around 4.6 GeV from $e^+e^-$ annihilation higher charmonia?,''
Phys. Rev. D \textbf{101} (2020) no.3, 034001
doi:10.1103/PhysRevD.101.034001
[arXiv:2001.00175 [hep-ph]].
%24 citations counted in INSPIRE as of 14 Nov 2023
\bibitem{Chen:2017abq}
D.~Y.~Chen, C.~J.~Xiao and J.~He,
%``Hidden-charm decays of Y(4390) in a hadronic molecular scenario,''
Phys. Rev. D \textbf{96} (2017) no.5, 054017
doi:10.1103/PhysRevD.96.054017
%12 citations counted in INSPIRE as of 14 Nov 2023
\bibitem{He:2019csk}
J.~He, Y.~Liu, J.~T.~Zhu and D.~Y.~Chen,
%``Y(4626) as a molecular state from interaction ${D}^*_s{\bar{D}}_{s1}(2536)-{D}_s{\bar{D}}_{s1}(2536)$,''
Eur. Phys. J. C \textbf{80} (2020) no.3, 246
doi:10.1140/epjc/s10052-020-7820-2
[arXiv:1912.08420 [hep-ph]].
%18 citations counted in INSPIRE as of 14 Nov 2023

%\cite{Peng:2022nrj}
\bibitem{Peng:2022nrj}
F.~Z.~Peng, M.~J.~Yan, M.~S\'anchez S\'anchez and M.~Pavon Valderrama,
%``Light- and heavy-quark symmetries and the Y(4230), Y(4360), Y(4500), Y(4620), and X(4630) resonances,''
Phys. Rev. D \textbf{107} (2023) no.1, 016001
doi:10.1103/PhysRevD.107.016001
[arXiv:2205.13590 [hep-ph]].
%10 citations counted in INSPIRE as of 03 Aug 2024

\bibitem{Liu:2008fh}
Y.~R.~Liu, X.~Liu, W.~Z.~Deng and S.~L.~Zhu,
%``Is $X(3872) $ Really a Molecular State?,''
Eur. Phys. J. C \textbf{56} (2008), 63-73
doi:10.1140/epjc/s10052-008-0640-4
[arXiv:0801.3540 [hep-ph]].
%168 citations counted in INSPIRE as of 14 Nov 2023


\bibitem{Wang:2020dya}
F.~L.~Wang and X.~Liu,
%``Exotic double-charm molecular states with hidden or open strangeness and around $4.5\sim 4.7$ GeV,''
Phys. Rev. D \textbf{102} (2020) no.9, 094006
doi:10.1103/PhysRevD.102.094006
[arXiv:2008.13484 [hep-ph]].
%14 citations counted in INSPIRE as of 14 Nov 2023
\bibitem{Liu:2009qhy}
X.~Liu, Z.~G.~Luo, Y.~R.~Liu and S.~L.~Zhu,
%``X(3872) and Other Possible Heavy Molecular States,''
Eur. Phys. J. C \textbf{61} (2009), 411-428
doi:10.1140/epjc/s10052-009-1020-4
[arXiv:0808.0073 [hep-ph]].
%218 citations counted in INSPIRE as of 14 Nov 2023
\bibitem{Sun:2012sy}
Z.~F.~Sun, X.~Liu, M.~Nielsen and S.~L.~Zhu,
%``Hadronic molecules with both open charm and bottom,''
Phys. Rev. D \textbf{85} (2012), 094008
doi:10.1103/PhysRevD.85.094008
[arXiv:1203.1090 [hep-ph]].
%31 citations counted in INSPIRE as of 14 Nov 2023

\bibitem{Faessler:2007gv}
A.~Faessler, T.~Gutsche, V.~E.~Lyubovitskij and Y.~L.~Ma,
%``Strong and radiative decays of the D(s0)*(2317) meson in the DK-molecule picture,''
Phys. Rev. D \textbf{76} (2007), 014005
doi:10.1103/PhysRevD.76.014005
[arXiv:0705.0254 [hep-ph]].
%193 citations counted in INSPIRE as of 05 Feb 2023
\bibitem{Faessler:2007us}
A.~Faessler, T.~Gutsche, V.~E.~Lyubovitskij and Y.~L.~Ma,
%``D* K molecular structure of the D(s1)(2460) meson,''
Phys. Rev. D \textbf{76} (2007), 114008
doi:10.1103/PhysRevD.76.114008
[arXiv:0709.3946 [hep-ph]].
%107 citations counted in INSPIRE as of 05 Feb 2023
\bibitem{Xiao:2020alj}
C.~J.~Xiao, Y.~B.~Dong, T.~Gutsche, V.~E.~Lyubovitskij and D.~Y.~Chen,
%``Towards the decay properties of deuteron-like state dN\ensuremath{\Omega},''
Phys. Rev. D \textbf{101} (2020), 114032
doi:10.1103/PhysRevD.101.114032
[arXiv:2004.12415 [hep-ph]].
%7 citations counted in INSPIRE as of 05 Feb 2023
\bibitem{Xiao:2019mvs}
C.~J.~Xiao, Y.~Huang, Y.~B.~Dong, L.~S.~Geng and D.~Y.~Chen,
%``Exploring the molecular scenario of Pc(4312) , Pc(4440) , and Pc(4457),''
Phys. Rev. D \textbf{100} (2019) no.1, 014022
doi:10.1103/PhysRevD.100.014022
[arXiv:1904.00872 [hep-ph]].
%111 citations counted in INSPIRE as of 05 Feb 2023
\bibitem{Chen:2015igx}
D.~Y.~Chen and Y.~B.~Dong,
%``Radiative decays of the neutral $Z_c(3900)$,''
Phys. Rev. D \textbf{93} (2016) no.1, 014003
doi:10.1103/PhysRevD.93.014003
[arXiv:1510.00829 [hep-ph]].
%34 citations counted in INSPIRE as of 05 Feb 2023
\bibitem{Xiao:2016hoa}
c.~J.~Xiao, D.~Y.~Chen and Y.~L.~Ma,
%``Radiative and pionic transitions from the $D_{s1}(2460)$ to the $D_{s0}^\ast(2317)$,''
Phys. Rev. D \textbf{93} (2016) no.9, 094011
doi:10.1103/PhysRevD.93.094011
[arXiv:1601.06399 [hep-ph]].
%22 citations counted in INSPIRE as of 05 Feb 2023
\bibitem{Weinberg:1962hj}
S.~Weinberg,
%``Elementary particle theory of composite particles,''
Phys. Rev. \textbf{130} (1963), 776-783
doi:10.1103/PhysRev.130.776
%588 citations counted in INSPIRE as of 05 Feb 2023
\bibitem{Salam:1962ap}
A.~Salam,
%``Lagrangian theory of composite particles,''
Nuovo Cim. \textbf{25} (1962), 224-227
doi:10.1007/BF02733330
%266 citations counted in INSPIRE as of 05 Feb 2023
\bibitem{vanKolck:2022lqz}
U.~van Kolck,
%``Weinberg's Compositeness,''
Symmetry \textbf{14} (2022), 1884
doi:10.3390/sym14091884
[arXiv:2209.08432 [hep-ph]].
%0 citations counted in INSPIRE as of 05 Feb 2023
\bibitem{Xiao:2020ltm}
C.~J.~Xiao, D.~Y.~Chen, Y.~B.~Dong and G.~W.~Meng,
%``Study of the decays of $S-$wave $\bar D^\ast K^\ast$ hadronic molecules: The scalar $X_0(2900)$ and its spin partners $X_{J(J=1,2)}$,''
Phys. Rev. D \textbf{103} (2021) no.3, 034004
doi:10.1103/PhysRevD.103.034004
[arXiv:2009.14538 [hep-ph]].
%40 citations counted in INSPIRE as of 14 Nov 2023

\bibitem{Casalbuoni:1996pg}
R.~Casalbuoni, A.~Deandrea, N.~Di Bartolomeo, R.~Gatto, F.~Feruglio and G.~Nardulli,
%``Phenomenology of heavy meson chiral Lagrangians,''
Phys. Rept. \textbf{281} (1997), 145-238
doi:10.1016/S0370-1573(96)00027-0
[arXiv:hep-ph/9605342 [hep-ph]].
%644 citations counted in INSPIRE as of 05 Feb 2023
\bibitem{Cheng:1992xi}
H.~Y.~Cheng, C.~Y.~Cheung, G.~L.~Lin, Y.~C.~Lin, T.~M.~Yan and H.~L.~Yu,
%``Chiral Lagrangians for radiative decays of heavy hadrons,''
Phys. Rev. D \textbf{47} (1993), 1030-1042
doi:10.1103/PhysRevD.47.1030
[arXiv:hep-ph/9209262 [hep-ph]].
%166 citations counted in INSPIRE as of 05 Feb 2023
\bibitem{Yan:1992gz}
T.~M.~Yan, H.~Y.~Cheng, C.~Y.~Cheung, G.~L.~Lin, Y.~C.~Lin and H.~L.~Yu,
%``Heavy quark symmetry and chiral dynamics,''
Phys. Rev. D \textbf{46} (1992), 1148-1164
[erratum: Phys. Rev. D \textbf{55} (1997), 5851]
doi:10.1103/PhysRevD.46.1148
%740 citations counted in INSPIRE as of 05 Feb 2023


\bibitem{Wise:1992hn}
M.~B.~Wise,
%``Chiral perturbation theory for hadrons containing a heavy quark,''
Phys. Rev. D \textbf{45} (1992) no.7, R2188
doi:10.1103/PhysRevD.45.R2188
%841 citations counted in INSPIRE as of 05 Feb 2023

\bibitem{Huang:2021kfm}
Q.~Huang, J.~Z.~Wang, R.~G.~Ping and X.~Liu,
%``Detecting the polarization in $\chi_{cJ} \to \phi \phi $ decays to probe hadronic loop effect,''
Phys. Rev. D \textbf{103} (2021) no.9, 096006
doi:10.1103/PhysRevD.103.096006
[arXiv:2102.07104 [hep-ph]].
%4 citations counted in INSPIRE as of 14 Nov 2023

\bibitem{Wang:2022jxj}
J.~Z.~Wang and X.~Liu,
%``Confirming the existence of a new higher charmonium \ensuremath{\psi}(4500) by the newly released data of e+e-\textrightarrow{}K+K-J/\ensuremath{\psi},''
Phys. Rev. D \textbf{107} (2023) no.5, 054016
doi:10.1103/PhysRevD.107.054016
[arXiv:2212.13512 [hep-ph]].
%5 citations counted in INSPIRE as of 14 Nov 2023
\bibitem{MARK-III:1988crp}
D.~Coffman \textit{et al.} [MARK-III],
%``Measurements of $J/\psi$ Decays Into a Vector and a Pseudoscalar Meson,''
Phys. Rev. D \textbf{38} (1988), 2695
[erratum: Phys. Rev. D \textbf{40} (1989), 3788]
doi:10.1103/PhysRevD.38.2695
%120 citations counted in INSPIRE as of 05 Feb 2023
\bibitem{DM2:1988bfq}
J.~Jousset \textit{et al.} [DM2],
%``The $J/\psi \to$ Vector + Pseudoscalar Decays and the $\eta$, $\eta^\prime$ Quark Content,''
Phys. Rev. D \textbf{41} (1990), 1389
doi:10.1103/PhysRevD.41.1389
%125 citations counted in INSPIRE as of 05 Feb 2023
\bibitem{Tornqvist:1993ng}
N.~A.~Tornqvist,
%``From the deuteron to deusons, an analysis of deuteron - like meson meson bound states,''
Z. Phys. C \textbf{61} (1994), 525-537
doi:10.1007/BF01413192
[arXiv:hep-ph/9310247 [hep-ph]].
%498 citations counted in INSPIRE as of 14 Nov 2023

%%%==================================Introduction==========================================
\bibitem{Wu:2021udi}
Q.~Wu, D.~Y.~Chen and T.~Matsuki,
%``A phenomenological analysis on isospin-violating decay of $X(3872)$,''
Eur. Phys. J. C \textbf{81} (2021) no.2, 193
doi:10.1140/epjc/s10052-021-08984-2
[arXiv:2102.08637 [hep-ph]].
%15 citations counted in INSPIRE as of 05 Feb 2023




\bibitem{Wu:2021caw}
Q.~Wu, D.~Y.~Chen and R.~Ji,
%``Production of $P_{cs}(4459)$ from $\Xi_b$ Decay,''
Chin. Phys. Lett. \textbf{38} (2021) no.7, 071301
doi:10.1088/0256-307X/38/7/071301
[arXiv:2103.05257 [hep-ph]].
%27 citations counted in INSPIRE as of 14 Nov 2023
\bibitem{Wu:2023fyh}
Q.~Wu, Y.~K.~Chen, G.~Li, S.~D.~Liu and D.~Y.~Chen,
%``Hunting for the hidden-charm molecular states with strange quarks in B and Bs decays,''
Phys. Rev. D \textbf{107} (2023) no.5, 054044
doi:10.1103/PhysRevD.107.054044
[arXiv:2302.01696 [hep-ph]].
%4 citations counted in INSPIRE as of 14 Nov 2023
\bibitem{Cheng:2004ru}
H.~Y.~Cheng, C.~K.~Chua and A.~Soni,
%``Final state interactions in hadronic B decays,''
Phys. Rev. D \textbf{71} (2005), 014030
doi:10.1103/PhysRevD.71.014030
[arXiv:hep-ph/0409317 [hep-ph]].
%417 citations counted in INSPIRE as of 14 Nov 2023
\bibitem{Gortchakov:1995im}
O.~Gortchakov, M.~P.~Locher, V.~E.~Markushin and S.~von Rotz,
%``Two meson doorway calculation for anti-p p ---\ensuremath{>} phi pi including off-shell effects and the OZI rule,''
Z. Phys. A \textbf{353} (1996), 447-453
doi:10.1007/BF01285155
%46 citations counted in INSPIRE as of 14 Nov 2023
\bibitem{Colangelo:2002mj}
P.~Colangelo, F.~De Fazio and T.~N.~Pham,
%``B- ---\ensuremath{>} K- (chi(c0)) decay from charmed meson rescattering,''
Phys. Lett. B \textbf{542} (2002), 71-79
doi:10.1016/S0370-2693(02)02306-7
[arXiv:hep-ph/0207061 [hep-ph]].
%122 citations counted in INSPIRE as of 05 Feb 2023

%\cite{BESIII:2023tll}
\bibitem{BESIII:2023tll}
M.~Ablikim \textit{et al.} [BESIII],
%``Measurement of e+e-\textrightarrow{}\ensuremath{\eta}J/\ensuremath{\psi} cross section from s=3.808\,\,GeV to 4.951~GeV,''
Phys. Rev. D \textbf{109} (2024) no.9, 092012
doi:10.1103/PhysRevD.109.092012
[arXiv:2310.03361 [hep-ex]].
%4 citations counted in INSPIRE as of 13 Aug 2024
\end{thebibliography}
\end{document}